\documentclass[aps,pra,twocolumn,groupedaddress,showpacs,showkeys]{revtex4}

\usepackage{graphicx}
\begin{document}

\newcommand{\be}{\begin{equation}}
\newcommand{\ee}{\end{equation}}

\title{Casimir effect in a superconducting cavity and the thermal controversy}

\author{ Giuseppe Bimonte}
\email[Bimonte@na.infn.it]
\affiliation{Kavli Institute for Theoretical Physics University of California Santa Barbara CA 93106-4030 USA\\  Dipartimento di Scienze Fisiche Universit\`{a} di
Napoli Federico II Complesso Universitario MSA, Via Cintia
I-80126 Napoli Italy and INFN Sezione di Napoli, ITALY\\
}

\date{\today}

\begin{abstract}
One of the most important and still unresolved problems in  the
field of dispersion forces, is that of determining the influence
of temperature on the Casimir force between two metallic plates.
While alternative theoretical approaches lead  to contradictory
predictions for the magnitude of the effect, no experiment has yet
detected the thermal correction to the Casimir force. In this
paper we show that a superconducting cavity  provides a new
opportunity to investigate the problem of the thermal dependence
of the Casimir force in real materials, by  looking at the change
of the Casimir force determined by a small change of temperature.
The actual feasibility of the proposed scheme is briefly
discussed.

\end{abstract}

\pacs{03.70.+k, 12.20.Ds, 42.50.Lc}
\keywords{Casimir,  thermal, superconducting,  proximity effects.}

\maketitle

\section{Introduction}

Dispersion forces have been the subject of intense theoretical and
experimental investigations for a long time. This is not
surprising, because these weak intermolecular forces have a truly
pervasive impact, from biology to chemistry, from physics to
engineering \cite{parse}. It may therefore come as a surprise to
know that there still exist, in this well established field,
unresolved problems of a fundamental character. This is indeed the
case with respect to the problem of determining the van der
Waals-Casimir interaction between two metallic bodies at finite
temperature \cite{bordag,brevik,qfext}. As of now, people simply
don't know how to compute it, and the numerous recent literature
on this subject provides contradictory recipes, which give widely
different predictions for its magnitude
\cite{bordag,brevik,qfext}. Apart from its intrinsic interest as a
problem in the theory of dispersion forces, addressing this
problem is important because many experiments on non-newtonian
forces at the submicron scale use metallic surfaces at room
temperature \cite{decca}. We remark that, while the Casimir
pressure has now been measured accurately in a number of
experiments \cite{lamor,decca}, there is at present only one
experiment that detected  the influence of temperature on the
Casimir-Polder interaction of a Bose-Einstein condensate with a
dielectric substrate \cite{cornell}. There are however no
experiments so far that detected the   temperature correction to
the Casimir force between two metallic bodies, and indeed there
are at present several ongoing and planned experiments to measure
it \cite{brown}. Since the theoretical solution  of this difficult
problem is still far, it is of the greatest importance to do
experiments to probe the effect. A recent accurate experiment
using a micromechanical oscillator \cite{decca} seems to favor one
of the theoretical approaches that have been proposed, but this
claim is disputed by other researchers \cite{brevik}.

In this paper, we propose a new experiment using superconducting
Casimir devices, that  could help to find an answer to the
question. Superconducting cavities are valuable tools to explore
important aspects of Casimir physics. Some time ago we proposed to
use {\it rigid} superconducting cavities to obtain the first
direct measurement of the change of Casimir energy that
accompanies the superconducting transition \cite{bimonte}. Here,
we argue that it  might be possible to shed some light on the
controversial problem of the thermal   Casimir  effect in metallic
systems, by measuring how the magnitude of the Casimir force {\it
changes}  as an effect of a change of temperature of the
superconducting apparatus. Indeed we show that certain
extrapolations to the superconducting state of existing
alternative approaches to the thermal Casimir effect for normal
metals, that have appeared in the recent literature, lead to
strikingly different predictions for the magnitude of the effect.
A distinctive feature of the proposed scheme is that, being a {\it
difference} force measurement, it might achieve a better
sensitivity than {\it absolute} measurements of the Casimir force.
The idea of exploiting difference force measurements to probe the
features of the thermal Casimir effect is not new indeed. Already
in Ref.\cite{chen} the possibility of observing the difference in
the thermal Casimir force in a {\it normal} (i.e. non
superconducting) metallic system, at two different temperatures,
was considered, opening up a new opportunity for the investigation
of the thermal Casimir effect in the submicron separation range.

The plan of the paper is a follows: in Sec. II we briefly review
the  thermal Casimir effect, and explain in some detail the
theoretical problems that arise in the case of metallic cavities.
In Sec. III we  describe our superconducting Casimir devices and
explain the general ideas on which our proposed experiment is
based. In Sec IV we discuss the important issue of the possible
prescriptions for the reflection coefficient of  quasi-static
magnetic fields by a superconducting plate, and in Sec V we apply
these prescriptions to the computation of the predicted change in
the Casimir force, in a plane-parallel system, as the temperature
of the superconductor is varied. In Sec VI we discuss the
experimentally important plate-sphere geometry, and finally in
Sec. VII, we present our conclusions and briefly discuss the
prospects of the proposed scheme of being experimentally feasible.

\section{The thermal Casimir effect in metallic cavities}

All current approaches to the thermal Casimir effect are based on
the theory of the Casimir effect developed long ago by Lifshitz
\cite{lifs}, on the basis of Rytov's general theory of e.m.
fluctuations \cite{rytov}. According to this theory, the Casimir
pressure $P$ between two plane-parallel (possibly stratified)
slabs at temperature $T$, separated by an empty gap of width $a$,
is: \be P(a,T)= \frac{k_B T}{2 \pi^2} \sum_{l \ge 0}{\,'} \int
\!\! d^2{\bf k_{\perp}} q_l\!\!\!\! \sum_{\alpha={\rm TE,TM}}
\left(\frac{e^{2 a q_l}}{r_{\alpha}^{(1)} r_{\alpha}^{(2)}} -1
\right)^{-1},\label{lifs} \ee where the plus sign corresponds to
an attraction between the plates. In this Equation,  the prime
over the $l$-sum means that the $l=0$ term has to taken with a
weight one half, $T$ is the temperature, ${\bf k_{\perp}}$ denotes
the projection of the wave-vector onto the plane of the plates and
$q_l =\sqrt{k_{\perp}^2+\xi_l^2/c^2}$, where $\xi_l= 2 \pi k_B T
l/\hbar$ are the Matsubara frequencies. The quantities
$r_{\alpha}^{(j)}\equiv r_{\alpha}^{(j)}(i \xi_l,{\bf
k_{\perp}})$, with $j=1,2$, denote the reflection coefficients for
the two slabs, for $\alpha$-polarization (for simplicity, we do
not consider the possibility of a non-diagonal reflection matrix),
evaluated at imaginary frequencies $i \xi_l$. It is of the outmost
importance at this point to make a remark on the range of validity
of Eq. (\ref{lifs}): Lifshitz original derivation assumed that
spatial dispersion is absent, so that the slabs can be modeled by
permittivities $\epsilon^{(j)}(\omega)$ and $\mu^{(j)}(\omega)$
depending only on the frequency $\omega$, and the reflection
coefficients $r_{\alpha}^{(j)}(i \xi_l,{\bf k_{\perp}})$ have the
familiar Fresnel expression. Nowadays it is well known
\cite{villareal} that Eq.(\ref{lifs}) is actually valid also for
spatially dispersive media (separated by an empty gap), like
metals at low temperature and obviously superconductors, provided
only that the appropriate reflection coefficients are used.

We now come to the main topic of the present paper, i.e. the
application of Lifshitz formula to non-magnetic metals (we set
$\mu$=1 from now on), which is the experimentally important case.
We suppose for simplicity that the plates are made of the same metal.
For our purposes, it is convenient first to write Eq. (\ref{lifs})
as a sum of three terms, ${P}_0^{(\rm TE)}$, ${P}_0^{(\rm TM)}$ and
${P}_1$, which include, respectively, the TE zero mode, the TM zero mode
and all non-vanishing Matsubara modes:
\be
{P}(a,T)={P}_0^{(\rm TE)}(a,T)+{P}_0^{(\rm TM)}(a,T)+{P_1}(a,T).\label{lifstwo}
\ee
We consider first $P_1(a,T)$. This term poses no particular problems.
At room temperature (300 K), the frequency $\xi_1$ of the first
non-vanishing Matsubara mode that enters in Eq. (\ref{lifs}) is
around $\xi_1=0.159 \,{\rm eV}/\hbar$, and belongs to the IR
region of the spectrum. On the other hand, the presence of the
exponential factor $\exp (2\,a q_l)$ cuts off the Matsubara modes
with frequencies greater than a few times the characteristic
cavity (angular) frequency $\omega_c=c/(2 a)$. For the relevant
experimental separations in the range from 100 nm to one or two
microns, the frequency $\omega_c$ falls somewhere in the range
from the near UV to the IR. In the frequency range extending from
$\xi_1$ to a few times $\omega_c$, spatial dispersion is negligible and one
can use the Fresnel formula  for the reflection coefficients, in
terms of the permittivity $\epsilon(i \xi_n)$. The values of
$\epsilon(i \xi_n)$ can be obtained, using dispersion theory, from
optical data of the plates (see however the footnote before Eq.
(\ref{drmpl}) below), or alternatively can be computed using
analytical models, like for example the plasma model of IR optics:\be
\epsilon_P(\omega)=1-\Omega_P^2/\omega^2\;,\label{plasma}\ee where
$\Omega_P$ is the plasma frequency. If needed, one can possibly augment
the above simple model by oscillator contributions accounting for
interband transitions that may become sizable at the higher
frequencies \cite{decca}. If desired, one can also take account of dissipation,
via the inclusion of a suitable set of relaxation frequencies in the
various terms. For the experimentally relevant separations $a$,
the inclusion of all these corrections changes a bit the magnitude of
$P_1(a,T)$, and poses no problems at all.

We consider now the two terms in Eq. (\ref{lifstwo}) that have $l=0$, namely
${P}_0^{(\rm TE)}(a,T)$ and ${P}_0^{(\rm TM)}(a,T)$
(we shall refer to these terms as zero modes).
Obviously, a vanishing complex imaginary frequency is the same as
a real vanishing frequency, and therefore the zero modes involve
the reflection coefficients of the slabs for static e.m. fields.
Let us examine first ${P}_0^{(\rm TM)}(a,T)$. A zero-frequency
${\rm TM}$ mode represents a static {\it electric} field. Since a
metal expels an electrostatic field, we clearly have $r_{\rm
TM}(0,{\bf k_{\perp}})=1$. Inserting this value of $r_{\rm TM}$
into the $l=0$ term of Eq. (\ref{lifs}) we see that in metals $P^{(\rm
TM)}_{0}(T)$
attains its maximum possible   value \be P^{(\rm
TM)}_{0}(a,T)=\frac{k_B \,T}{8 \pi a^3}\,\zeta(3) \,,\label{TMzm}\ee where $\zeta(3)\simeq1.20$ is
Riemann zeta function.

Consider now the ${\rm TE}$ zero mode $P^{(\rm
TE)}_{0}(a,T)$. Surprisingly, this single
term has become the object of a  controversy, in the last
ten years, and no consensus has been reached yet about its correct
value. The whole crisis of the thermal Casimir effect for metallic
bodies in fact stems from this very term, and it is therefore
useful to review briefly what the problem is. A zero-frequency
${\rm TE}$ mode represents a static magnetic field. Since non
magnetic metals are transparent to magnetic fields, it seems
natural to suppose that we should take \be r_{\rm TE}(0,{\bf
k_{\perp}})\vert_{\rm Dr}=0. \label{drumod}\ee When this value is
inserted in Eq. (\ref{lifs}) one finds easily that the ${\rm TE}$
zero mode gives a vanishing contribution to the Casimir pressure:
\be
P_0^{\rm (TE)}\vert_{\rm Dr}=0\,.\label{Pdr}
\ee
This result is dubbed in the Casimir community as the Drude-model
prediction \cite{sernelius}, because $r_{\rm TE}(0,{\bf
k_{\perp}})=0$ is the value that is obtained by inserting into the
Fresnel formula for the ${\rm TE}$ reflection coefficient the
familiar Drude extrapolation of the permittivity of a metal to low
frequencies: \be \epsilon_D(\omega)=1-\Omega_P^2/[\omega(\omega+i
\gamma)]\;.\label{drude}\ee In this formula,  $\gamma$ is the
relaxation frequency accounting for ohmic conductivity. Despite
its physical plausibility, this Drude approach has been much
criticized in the recent literature. The first problem that was
encountered is theoretical, as it was shown that, in the case of
perfect lattices, the Drude model value of the ${\rm TE}$ zero
mode is apparently inconsistent with Nernst theorem
\cite{bezerra}. The problem is subtle though, and the actual
existence of this inconsistency
is still disputed. For a thorough discussion
 of different points of view on this issue we refer the reader to the recent
 review \cite{brevik} (see also Ref.\cite{bimo3}). In addition,
 and perhaps more importantly, it has
been claimed that the Drude model is also inconsistent with
recent experiments \cite{decca}. The solution that has been
proposed \cite{decca} to cure both difficulties is very puzzling:
in evaluating the low-frequency contributions of Eq. (\ref{lifs}),
and in particular the  ${\rm TE}$ zero mode, instead of the
physically plausible Drude model, one should use the {\it plasma
model} of IR optics, Eq.(\ref{plasma}), extrapolated {\it without
modifications} all the way to zero frequency.  For the ${\rm TE}$
zero mode, this prescription leads to the following expression for
the reflection coefficient: \be r_{\rm TE}(0,{\bf
k_{\perp}})\vert_{\rm pl}= \frac{
\sqrt{\Omega_P^2/c^2+k_{\perp}^2}-k_{\perp}}{
\sqrt{\Omega_P^2/c^2+k_{\perp}^2}+k_{\perp}}\;, \label{plamod}\ee
and we shall refer to this as the plasma model prescription. For
the relevant values of $k_{\perp}\simeq 1/(2a)$, and for typical
values of $\Omega_P\simeq 9\,{\rm eV}/\hbar$ (corresponding to
gold), it is easy to verify that $r_{\rm TE}(0,{\bf
k_{\perp}})\vert_{\rm pl}$ is close to one. For example, for
$a=200$ nm, we obtain $r_{\rm TE}(0,{1/(2a)})\vert_{\rm pl}=0.90$.
Since $r_{\rm TE}(0,{\bf k_{\perp}})\vert_{\rm pl}\neq 0$, $P_0^{\rm
(TE)}(a,T)$ does not vanish anymore, and we let $P_0^{\rm
(TE)}(a,T;\Omega_P)\vert_{\rm pl}$ its magnitude, as determined by the plasma prescription.
For  $\omega_c/\Omega_P \ll 1$,  $P_0^{\rm
(TE)}(a,T;\Omega_P)\vert_{\rm pl}$ has the following expansion: \be
P_0^{\rm (TE)}(a,T;\Omega_P )\vert_{\rm pl} \simeq \frac{k_B \,T}{8 \pi a^3}\,\zeta(3)\left(1-6 \frac{\delta }
{a} + 24 \frac{\delta^2 }{a^2} \right)\,,\label{Pzero}\ee where $\delta=c/\Omega_P$.
We see that, for sufficiently large separations $a$, the magnitude of $P_0^{\rm
(TE)}(a,T;\Omega_P)\vert_{\rm pl}$ is comparable to that  $P_0^{\rm
(TM)}(a,T)$. Recalling
that the Drude prescription predicts instead $P_0^{\rm
(TE)}\vert_{\rm Dr}=0$, we see that the predicted Casimir pressures for the two
prescriptions differ by a quantity $\widetilde{\Delta P}(a,T) \equiv P(a,T)\vert_{\rm
Dr}-P(a,T)\vert _{\rm pl}$ \footnote{The actual difference $\widetilde{\Delta
P}(a,T)$ between the predicted thermal Casimir pressures is slightly
different from the value quoted in Eq.(\ref{drmpl}). The reason for this is that a low
frequency extrapolation of the optical data is required also for
the evaluation, by means of dispersion theory, of $\epsilon(i
\xi_l)$ for $l>0$, and therefore the two prescriptions predict
slightly different magnitudes also for $P_1(a,T)$ (see Ref. \cite{decca} for details).}:
\be \widetilde{\Delta P}(a,T) =  -P_0^{\rm (TE)}(a,T;\Omega_P)\vert_{\rm pl}
\,.\label{drmpl}\ee The minus sign in the r.h.s. means that the
Drude model predicts a {\it smaller} Casimir pressure than the
plasma model.

The disagreement between the two prescriptions is most evident at large
separations. Indeed, for $k_B T a/(\hbar c) \gg 1$, $P_1(a,T)$ becomes negligible and the entire
Casimir pressure is given by the zero modes contribution only \cite{bordag, brevik}.
Since in this limit $P_0^{\rm (TE)}(a,T;\Omega_P)\vert_{\rm pl} \simeq P^{(\rm
TM)}_{0}(T)$, we see at once that the pressure predicted by the Drude prescription is
one-half that predicted by the plasma model prescription!
Obviously then the straightforward way to clarify the problem would be to
measure the Casimir pressure at large separations, say around 5 $\mu$m, for room temperature.
Unfortunately, at such large separations the Casimir pressure is very small, and as a result
all attempts made so far to measure it have been unsuccessful.

At the time of this writing, the most significant experiments are
the ones reported in \cite{decca}, in which the Casimir pressure
was measured for separations in the range from 160 to 750 nm. For
such small separations, the quantity $\widetilde{\Delta P}(a,T)$
is only a small fraction of the absolute Casimir pressure (for
gold plates at room temperature, about two-three percent at
separations of around 200-300 nm, where the experiment in the
second of Refs.\cite{decca} was most sensitive). Observing such a
small correction via an absolute Casimir measurement is very
difficult, because one needs to measure with high precision the
plates separation, and  control a number of systematic errors,
like residual electrostatic attractions \cite{patch}, roughness of
the plates etc. After an accurate analysis of these possible
sources of systematic errors, the authors of Ref.\cite{decca}
conclude that the Drude approach is inconsistent with the
experiment at high confidence level. However, this statement has
been disputed by other researchers (see the second of
Refs.\cite{bordag}). In this regard, it has been remarked recently
\cite{piro} that obtaining an accurate theoretical prediction,  at
the percent level, for the magnitude of Casimir force is barely
possible, unless one has accurate optical data, extending over a
wide range of frequencies around $\omega_c$, taken on the actual
surfaces used in the experiment.

\section{A superconducting Casimir experiment}

In the light of the above considerations, we wondered  if one can conceive a new type of experiment
specifically devised to probe the magnitude of the contribution from the ${\rm TE}$ zero mode.
As we  see below,   a superconducting Casimir apparatus  is  in principle a good tool to do that.

Consider then two metallic plates made of a superconducting metal,
and imagine cooling them at a temperature $T_1 < T_c$, where $T_c$
is the critical temperature. We suppose that the equilibrium
temperature of the system is rapidly changed from the temperature
$T_1$ to a higher temperature $T_2 < T_c$. It is our aim to
estimate the difference $ {\Delta P}$ between the Casimir
pressures at the two temperatures: \be{\Delta P}(a \vert
T_2,s;T_1,s)\equiv P(a \vert T_2,s)-P(a\vert
T_1,s)\,.\label{nota}\ee Let us explain the meaning of the new
notation used in the above Equation.  The notation is used to
remind the two-fold dependence of the Casimir pressure on the
temperature $T$. Indeed, by looking at Eq.(\ref{lifs}) we see
that, on one side, $P$ has an {\it explicit} temperature
dependence, determined by the overall $T$-factor, in front of the
sum symbol, and by the $T$-dependence of the Matsubara frequencies
$\xi_l=2 \pi k_B T l/\hbar$. More importantly for our purposes,
there is however also an {\it implicit} dependence of $P$ on $T$,
through the reflection coefficients, which may themselves depend
on the temperature. The notation in Eq.(\ref{nota}) stresses this
fact, by making explicit reference to the dependence of $P$ on the
state of the plates, which can either be normal ($n$) or
superconducting ($s$). We remark that probing the explicit
$T$-dependence of the Casimir pressure, by letting pretty large
changes of temperature (around 50 K), was the goal of the
room-temperature experiment proposed in Ref.\cite{chen}. In that
experiment, the temperature-dependence of the reflection
coefficients was completely negligible. In our case the situation
is much the opposite:  the main expected cause of variation of
$P(a,T)$ is now the  temperature dependence (in a certain
frequency range) of the reflection coefficients of the
 plates in the  superconducting state.

After these remarks, we can go now to the main point: why should the proposed experiment be capable
of telling us anything about the thermal Casimir problem in metallic systems? This is so, because
the superconducting transition affects
 the optical properties of a metal only at  frequencies $\omega$ smaller than a few times
 $k_B T_c/\hbar$ \cite{glover}.  As a result, the magnitude of ${\Delta P}$  turns out to be
 almost completely determined by the ${\rm TE}$
zero mode, and therefore it is very sensitive to the prescription adopted for it. We shall
indeed show below that the two prescriptions considered in the previous Section, lead to sharply
different magnitudes for $ {\Delta P}$.

Before we embark in the computation of $\Delta P$, a key problem to address is to decide
what prescription  are we going to use for the reflection coefficient of the TE zero mode in
the superconducting state. This is the subject of the next Section.

\section{The TE zero mode in a superconductor}

According to Lifshitz formula, which we recall is valid also for superconductors, in order to
estimate the contribution of the TE zero mode to the Casimir pressure for a superconducting cavity,
we need to determine what expression for the reflection coefficient $r_{\rm
TE}^{(s)}(0,{\bf k_{\perp}})$  should be used in Eq.(\ref{lifs}), when a plate is superconducting.
Addressing this problem is the purpose of this Section.

We have seen in  Section II that there is a theoretical
uncertainty in the value of the reflection coefficient $r_{\rm
TE}^{(n)}(0,{\bf k_{\perp}})$ of a metallic plate, in the normal
state, depending on whether we include or not, in the permittivity
of the metal, the effect of ohmic dissipation. If dissipation is
included, we end up with the Drude prescription,
Eq.(\ref{drumod}), if dissipation is ignored we instead obtain the
plasma prediction, Eq.(\ref{plamod}). What about a superconductor?
For sure, static currents flow in a superconductor without any
measurable dissipation, and so there should really be no room for
ambiguity now: for what we know, in the dc limit, superconductors
are strictly dissipationless. This is reflected in the plasma-like
form of the permittivity function $\epsilon_s(\omega)$ that can be
used to describe the response of superconductors to quasi-static
electromagnetic fields, in the local London limit: \be
\epsilon_s(\omega)=-[c/{(\lambda_L \omega)}]^2.\ee In this
Equation, $\lambda_L$ represents the penetration depth, that can
be expressed in terms of the plasma frequency $\Omega_P$ as
$\lambda_L=(n/n_s)^{1/2}c/\Omega_P$, where $n_s/n$ represents the
fractional density  of superconducting electrons. The temperature
dependence of  $n_s/n$ is rather well described by the
phenomenological law $n_s/n=1-(T/T_c)^4$ \cite{tinkam}. If we
insert $\epsilon_s(\omega)$ into the Fresnel formula for the TE
reflection coefficient at zero frequency  we obatin: \be r_{\rm
TE}^{(s)}(0,{\bf k_{\perp}}\vert T)= \frac{
\sqrt{1/\lambda_L^2+k_{\perp}^2}-k_{\perp}}{
\sqrt{1/\lambda_L^2+k_{\perp}^2}+k_{\perp}}\;,\;\;\;{\rm for}\;T
\le T_c \label{superTE}\ee where we stressed that $r_{\rm
TE}^{(s)}$ is temperature dependent, because $\lambda_L$ is. The
above form of $r_{\rm TE}^{(s)}(0,{\bf k_{\perp}})$ is physically
plausible, and it correctly describes the Meissner effect. It
should be noted that $r_{\rm TE}^{(s)}(0,{\bf k_{\perp}})$ has the
same form as the plasma-model prescription $r_{\rm
TE}^{(n)}(0,{\bf k_{\perp}})\vert_{\rm pl}$ for the normal state,
apart from the replacement of $\Omega_P^2/c^2$ by $1/\lambda_L^2$.
In fact, taking account of the temperature dependence of
$\lambda_L$, $r_{\rm TE}^{(s)}(0,{\bf k_{\perp}}\vert T)$ provides
a smooth interpolation between $r_{\rm TE}^{(n)}(0,{\bf
k_{\perp}})\vert_{\rm pl}$ and $r_{\rm TE}^{(n)}(0,{\bf
k_{\perp}})\vert_{\rm Dr}$,  as the temperature is increased from
zero towards $T_c$: \be r_{\rm TE}^{(s)}(0,{\bf k_{\perp}}\vert T)
\rightarrow r_{\rm TE}^{(n)}(0,{\bf k_{\perp}})\vert_{\rm pl}
\,\;\;\;{\rm for}\;\;T/T_c \rightarrow 0,\ee \be r_{\rm
TE}^{(s)}(0,{\bf k_{\perp}}\vert T) \rightarrow r_{\rm
TE}^{(n)}(0,{\bf k_{\perp}})\vert_{\rm Dr}=0 \,\;\;\;{\rm
for}\;\;T/T_c \rightarrow 1.\ee Now we are faced with the
following question: depending on what prescription we choose in
the normal state, what do we do in the superconducting state? We
cannot offer a certain answer to this question, but the following
choices appear to us as reasonable. Consider first the Drude
prescription. Since the superconducting phase transition is of
second order, the reflection coefficients must be smooth across
$T_c$. Now, the Drude prescription asserts that $r_{\rm
TE}^{(n)}(0,{\bf k_{\perp}})$ is zero, and therefore we should fix
the prescription in the superconducting phase in such a way that
$r_{\rm TE}^{(s)}(0,{\bf k_{\perp}}\vert T)$ approaches zero for
$T\rightarrow Tc$. Of course, we could achieve this by taking
$r_{\rm TE}^{(s)}(0,{\bf k_{\perp}}\vert T)$ identically zero for
$T<T_c$. However, such an ansatz looks to us physically wrong,
because it is in conflict with the Meissner effect. The
prescription that we choose then is to use Eq.(\ref{superTE}),
with $\lambda_L$ temperature dependent, as our Drude recipe in the
superconducting phase: \be r_{\rm TE}^{(s)}(0,{\bf k_{\perp}}\vert
T)\vert_{\rm Dr}=r_{\rm TE}^{(s)}(0,{\bf k_{\perp}}\vert
T)\,\,\;\;\;{\rm for}\;T \le T_c.\label{drsup} \ee This choice
ensures smoothness of the reflection coefficient at the phase
transition, while describing correctly the Meissner effect.
Importantly, within this prescription, the reflection coefficient
is {\it temperature dependent} in the superconducting phase.

Consider now the plasma prescription. Again, the reflection
coefficient must be smooth through the transition. If we further
make the plausible assumptions that the reflection coefficient
cannot decrease as we reduce the temperature, and that the plasma
frequency sets up an upper limit on the quantity $c/\lambda_L$, we
are led to conclude that, within the plasma prescription, the
reflection coefficient $r_{\rm TE}^{(s)}(0,{\bf k_{\perp}}\vert
T)$ has to be temperature independent and equal to its
normal-state value: \be r_{\rm TE}^{(s)}(0,{\bf k_{\perp}}\vert
T)\vert_{\rm pl}=r_{\rm TE}^{(n)}(0,{\bf k_{\perp}})\vert_{\rm
pl}\,\;\;\;{\rm for}\;T \le T_c.\label{plsup} \ee It should be
noted that, with this prescription, the reflection coefficient
does not change at all across the transition, and it is {\it
temperature independent} in the superconducting phase. This
feature marks a crucial difference between the plasma prescription
and our Drude prescription in Eq.(\ref{drsup}).

Having defined our prescriptions for the TE reflection coefficient in the
superconducting state, we are ready now to proceed with the computation of $\Delta P$ according
to the two approaches.

\section{Computing $\Delta P$}

In this Section, we undertake the computation of $\Delta P$.
We shall consider two possible geometries: the  plane-parallel configuration, and the sphere-plate
configuration, which is the one usually adopted in
current experiments. For either geometry,
it is useful to consider two different setups for the system: the first one consists of
two superconducting plates, while the second setup has just one superconducting plate, the other plate
 being made of a normal metal. To be definite, we shall consider Nb as our superconducting material,
 and Au as the normal metal.
The choice of Nb is
due to the fact that, among the classic metallic superconductors,
it is the one with the highest critical temperature $T_c=9.2$ K, a
feature that will be seen to ensure  the largest difference
between the Drude and the plasma predictions for $ {\Delta P}$. It is assumed that the Nb plates have a thickness
much larger than the penetration depth $\lambda_L$ of magnetic
fields.
In this Section we shall only consider the plane-parallel case. The sphere-plate case will be
treated in Sec.VI below.

It is convenient to split Eq.(\ref{nota}) in a way analogous to
Eq. (\ref{lifstwo}): \be {\Delta P}={\Delta P}_0^{(\rm
TE)}+{\Delta P}_0^{(\rm TM)}+{\Delta P_1},\label{splitone} \ee
where the meaning of the symbols is obvious. Leaving aside for the
moment ${\Delta P}_0^{(\rm TE)}$, we consider first ${\Delta
P}_0^{(\rm TM)}$ and ${\Delta P_1}$. Since $r_{\rm TM}(0,{\bf
k_{\perp}})$ is one for metals, no matter whether normal of
superconducting, we surely have: \be {\Delta P}_0^{(\rm TM)}(a
\,\vert T_2,s; T_1,s)={\Delta P}_0^{(\rm TM)}(a \,\vert T_2,n;
T_1,n)\;. \ee Consider now ${\Delta P_1}$. Recall that ${\Delta
P_1}$ includes only the contributions from non vanishing Matsubara
modes. The key experimental fact to consider now is that the
optical properties of superconductors are appreciably different
from those of the normal state (right above $T_c$) only for
frequencies less than a few times $k_B T_c/\hbar$ \cite{glover}.
Since $\xi_1(T_2) \simeq \xi_1(T_1)\simeq\xi_1(T_c)  = 2 \pi k_B
T_c/\hbar$, we see that already the first non-vanishing Matsubara
mode is   over six times larger than $k_B T_c/\hbar$. Therefore,
in the range of frequencies that is relevant for $\Delta P_1$, the
superconductor can be described, at all temperatures below $T_c$,
by the same set of reflection coefficients  of the normal state,
at temperatures slightly above $T_c$ (this consideration applies
even more to Au, of course). This implies at once that \be {\Delta
P}_1(a \,\vert T_2,s; T_1,s)={\Delta P}_1(a \,\vert T_2,n;
T_1,n)\;. \ee At this point, it is convenient to add and subtract
from Eq.(\ref{splitone}) the quantity ${\Delta P}_0^{(\rm TE)}(a
\,\vert T_2,n; T_1,n)$. In this way, we get
\begin{widetext}
\begin{eqnarray}
  {\Delta P}(a \,\vert T_2,s; T_1,s) &=& [{\Delta P}_0^{(\rm TE)}(a \,\vert T_2,s; T_1,s)-
{\Delta P}_0^{(\rm TE)}(a \,\vert T_2,n; T_1,n)] \nonumber \\
   &+&[{\Delta P}_0^{(\rm TE)}(a \,\vert T_2,n; T_1,n)+{\Delta P}_0^{(\rm TM)}
   (a \,\vert T_2,n; T_1,n)+{\Delta P}_1(a \,\vert T_2,n; T_1,n)].\label{splittwo}
\end{eqnarray}
 \end{widetext}
Now we note that the sum of three terms on the second line of the above Equation is nothing but
the quantity ${\Delta P}(a \,\vert T_2,n; T_1,n)$ that represents the change of Casimir pressure
when the temperature is changed from $T_1$ to $T_2$,
in a normal metal system, with temperature independent reflection coefficients.
This quantity has been computed already in Ref.\cite{chen}, both for the Drude and
the plasma prescriptions, and we shall later use the formulae derived in that paper to estimate it.
Substituting ${\Delta P}(a \,\vert T_2,n; T_1,n)$, we may then rewrite Eq.(\ref{splittwo}) as:
\begin{widetext}
\be {\Delta P}(a \,\vert T_2,s; T_1,s) = [{\Delta P}_0^{(\rm
TE)}(a \,\vert T_2,s; T_1,s)- {\Delta P}_0^{(\rm TE)}(a \,\vert
T_2,n; T_1,n)]+{\Delta P}(a \,\vert T_2,n;
T_1,n)\,.\label{keysplit} \ee
\end{widetext}
The above Equation represents the first key result of this paper,
because it shows, as announced earlier, that  superconductivity
has an effect on the Casimir pressure only via the TE zero mode.
At this point we consider separately the plasma and the Drude
prescriptions.

\subsection{The plasma prescription}

According to our plasma prescription for the superconducting
state, Eq.(\ref{plsup}), the reflection coefficient for the TE
zero mode in the superconducting state has precisely the same
expression as in the normal state. This being so, we obviously
have: \be {\Delta P}_0^{(\rm TE)}(a \,\vert T_2,s;
T_1,s)\vert_{\rm pl}- {\Delta P}_0^{(\rm TE)}(a \,\vert T_2,n;
T_1,n)\vert_{\rm pl}=0\,, \ee both for the Nb-Nb and the Nb-Au
setups. When this formula is used in Eq.(\ref{keysplit}), we see
that, in either setup,  superconductivity has no effect on $\Delta
P$: \be {\Delta P}(a \,\vert T_2,s; T_1,s)\vert_{\rm pl} = {\Delta
P}(a \,\vert T_2,n; T_1,n)\vert_{\rm pl}\,.\label{plasmaeff} \ee
The quantity ${\Delta P}(a \,\vert T_2,n; T_1,n)\vert_{\rm pl}$
has been computed in Ref.\cite{chen}, for the case of two plates
made of the same metal. This case is good enough for us, also in
the case of the Nb-Au setting, because of the small difference
between the plasma frequencies of Nb (8.7 eV/$\hbar$) and Au (9
eV/$\hbar$). In our computations, from now on, we shall than take
for both metals the common value $\Omega_P=9$ eV/$\hbar$. Then,
our ${\Delta P}(a \,\vert T_2,n; T_1,n)\vert_{\rm pl}$ coincides
with minus the quantity $\Delta F_{pp}$ of Ref.\cite{chen} (the
minus sign is due to the fact that we consider positive forces to
correspond to an attraction, while Ref.\cite{chen} uses the
opposite convention). Therefore, we can rewrite
Eq.(\ref{plasmaeff}) as \be {\Delta P}(a \,\vert T_2,s;
T_1,s)\vert_{\rm pl} =  -\Delta F_{pp}\,.\label{plasmapredic} \ee
We can obtain an estimate of the effect by using the following
perturbative expression for $\Delta F_{pp}$ of Ref.\cite{chen},
that holds for low temperatures $k_B T a/(\hbar c) \ll 1$, and for
not too small  separations $\omega_c/\Omega_p \ll 1$: \be \Delta
F_{pp}(a,T_1,T_2)=-\Delta^{(1)}F_{pp}(T_1,T_2)\,\Delta^{(2)}F_{pp}(a,T_1,T_2)\,,
\ee where \be \Delta^{(1)}F_{pp}(T_1,T_2)=\frac{\pi^2
k_B^4(T_2^4-T_1^4)}{45 \hbar^3 c^3}\,, \ee and \be
\Delta^{(2)}F_{pp}(a,T_1,T_2)=1+\frac{90 \zeta
(3)}{\pi^3}\frac{\delta}{a}\frac{T_{\rm eff}}{T_1+T_2}\left(
1+\frac{T_1 T_2}{T_1^2+T_2^2}\right), \ee where $k_B T_{\rm
eff}=\hbar c/(2a)$ and $\delta=c/\Omega_p$. The above formulae
show that ${\Delta P}({\rm Nb}-{\rm Nb})\vert_{\rm pl}$ and
${\Delta P}({\rm Nb}-{\rm Au})\vert_{\rm pl}$ are both extremely
small. For example, upon taking $T_1=5$ K, $T_2 \simeq T_c$, and
$a=100$ nm, we obtain: \be
 {\Delta P}\vert_{\rm pl} \approx 1.4\times10^{- 9}\,{\rm Pa}\;.
\ee The conclusion that we draw is that, with the plasma
prescription, the change with temperature of the Casimir pressure
is unmeasurably small, both in the Nb-Nb and in the Nb-Au setups.

\subsection{The Drude prescription}

 We now consider the Drude prescription. First, we analyze the Nb-Au setup.
 Since Au is always a normal metal, its Drude reflection coefficient for the TE zero mode is zero,
 at both temperatures. Therefore, since Lifshitz formula implies only products of reflection
 coefficients for  the two plates, it follows that the TE zero mode gives no contribution, in the Nb-Au setup:
 \begin{eqnarray}
   {\Delta P}_0^{(\rm TE)}(a \,\vert T_2,s; T_1,s)\vert_{\rm Dr}^{\rm Nb-Au} &=& \nonumber \\
   {\Delta P}_0^{(\rm TE)}(a \,\vert T_2,n; T_1,n)\vert_{\rm Dr}^{\rm Nb-Au} &=& 0.
 \end{eqnarray}
Upon substituting these formulae into Eq.(\ref{keysplit}) we
obtain: \be {\Delta P}(a \,\vert T_2,s; T_1,s)\vert_{\rm Dr}^{\rm
Nb-Au} = {\Delta P}(a \,\vert T_2,n; T_1,n)\vert_{\rm Dr}^{\rm
Nb-Au}\,.\label{DrudeNbAu} \ee Again, we find that
superconductivity has no effect on the change of Casimir pressure.
The Drude value of the quantity ${\Delta P}(a \,\vert T_2,n;
T_1,n)\vert_{\rm Dr}^{\rm Nb-Au}$ can also be found in
Ref.\cite{chen}, from which we obtain the estimate:
\begin{eqnarray}
  {\Delta P}(a \,\vert T_2,s; T_1,s)\vert_{\rm Dr}^{\rm Nb-Au}&=& -\Delta F_{pp} \nonumber \\
  -\frac{\zeta(3) k_B( T_2-T_1)}{8 \pi a^3}& & \!\!\!\!\!\!
 \left(1-6 \frac{\delta}{a}+24 \frac{\delta^2}{a^2} \right) .\label{DrNbAu}
\end{eqnarray}
 The expression on the right hand side of the above Equation, differs from the analogous
 expression obtained within the plasma prescription, Eq.(\ref{plasmapredic}) by the
 presence of the extra term on the second line. At our cryogenic temperatures, the
 absolute value of the latter term is larger than $\vert\Delta F_{pp}\vert$ by several
 orders of magnitude, for changes of temperature of a few degrees. We therefore see that
 heating the system, the Casimir pressure decreases by an amount proportional to the the
 temperature change. We remark once again that this change of Casimir pressure is not an
 effect of superconductivity, and it only arises from the
explicit $T$ dependence of the Lifshitz formula. Indeed, an analogous phenomenon was found to
occur in the room temperature setting of Ref.\cite{chen}.

We finally consider the Nb-Nb setup, the most interesting case
indeed. Using the reflection coefficient in Eq.(\ref{drsup}), we
easily find $$ {\Delta P}_0^{(\rm TE)}(a \,\vert T_2,s;
T_1,s)\vert_{\rm Dr}^{\rm Nb-Nb}=$$ \be P_0^{\rm (TE)}
(a,T_2;c/\lambda_L)-P_0^{\rm (TE)} (a,T_1;c/\lambda_L), \ee \be
{\Delta P}_0^{(\rm TE)}(a \,\vert T_2,n; T_1,n)\vert_{\rm Dr}^{\rm
Nb-Nb}=0, \ee where, in the first of the two above Equations,
 $P_0^{\rm (TE)}(a,T;c/\lambda_L)$ denotes the magnitude of the TE zero mode, that results after
we plug into the $l=0$ term of Lifshitz formula the expression the Drude ansatz for the TE reflection
coefficient for the superconductor, Eq.(\ref{drsup}).
As for the quantity, ${\Delta P}(a \,\vert T_2,n; T_1,n)\vert_{\rm Dr}$, its magnitude is
given by the same expression on the right hand side of Eq.(\ref{DrNbAu}).
Collecting everything together, Eq. (\ref{keysplit}) gives:
\begin{widetext}
\be {\Delta P}(a \,\vert T_2,s; T_1,s)\vert_{\rm Dr}^{\rm
Nb-Nb}=[P_0^{\rm (TE)}(a,T_2;c/\lambda_L)-P_0^{\rm
(TE)}(a,T_1;c/\lambda_L)] -\Delta F_{pp}
  -\frac{\zeta(3) k_B (T_2-T_1)}{8 \pi a^3}
 \left(1-6 \frac{\delta}{a}+24 \frac{\delta^2}{a^2} \right).\label{DrNbNb}
\ee
\end{widetext}
In Fig.1 we show plots of ${\Delta P}(a \,\vert T_2,s;
T_1,s)\vert_{\rm Dr}^{\rm Nb-Nb}$ (solid line) and ${\Delta P}(a
\,\vert T_2,s; T_1,s)\vert_{\rm Dr}^{\rm Nb-Au}$ (dashed line),
both expressed in mPa, for $a=$ 150 nm, as a function of the
temperature $T_1$ (in K), for $T_2 \simeq T_c$. In Fig.2 we show
the plots of the same quantities, this time as functions of plates
separation $a$ (in microns), for $T_1=5$ K and $T_2 \simeq T_c$.
Note that, contrary to what we found with the plasma prescription,
our Drude prescription predicts negative changes of pressure, i.e.
the Casimir pressure decreases as one goes from the
superconducting state towards the normal state. Moreover, it
should also be noted that the change of pressure is larger in the
Nb-Nb setup, than in the Nb-Au setup. However, the really striking
finding is that, for both setups, the Drude approach predicts
changes of Casimir pressure, whose magnitudes are around five or
six orders of magnitude larger than the corresponding changes
predicted by the plasma approach!

\begin{figure}
\includegraphics{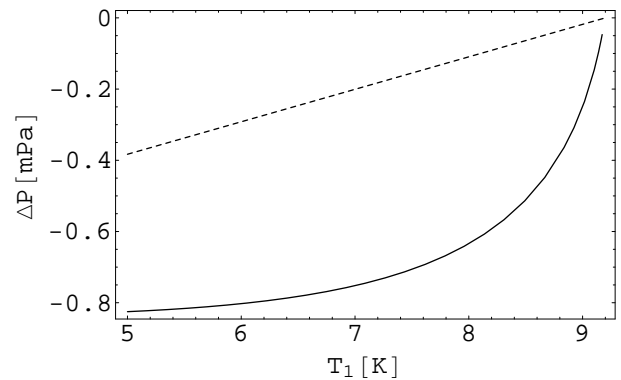}
\caption{\label{fig1} Plots of ${\Delta P}({\rm Nb}-{\rm
Nb})\vert_{\rm Dr}$ (solid line) and ${\Delta P}({\rm Nb}-{\rm
Au})\vert_{\rm Dr}$ (dashed line) (in mPa) for two parallel plates
at fixed separation $a=$ 150 nm, versus temperature $T_1$ (in K),
for fixed $T_2\simeq T_c$.}
\end{figure}

\begin{figure}
\includegraphics{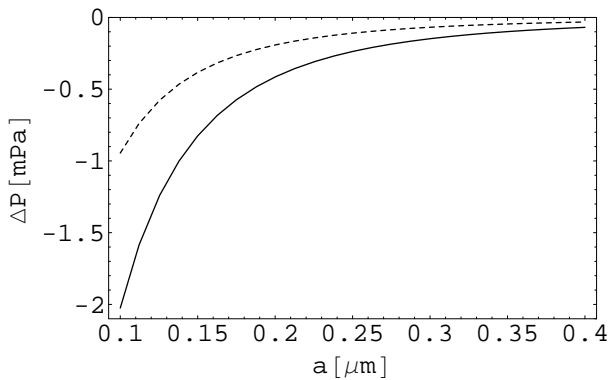}
\caption{\label{fig2} Plots of ${\Delta P}({\rm Nb}-{\rm
Nb})\vert_{\rm Dr}$ (solid line) and ${\Delta P}({\rm Nb}-{\rm
Au})\vert_{\rm Dr}$ (dashed line) (in mPa) for two parallel plates
versus separation $a$ (in microns), for  $T_1=5$ K  and $T_2
\simeq T_c$.}
\end{figure}

\section{The sphere-plate case}

Now we consider the configuration of a sphere  placed above a plate, which is the one used in most of the current experiments.
We let $R$ the radius of the sphere
and $a$ the minimum sphere-plate separation. As it is well known, for small
separations, i.e. for $a/R \ll 1$, the Casimir force $F_{\rm sp}$  between a sphere and a plate can
be obtained by means of the so called proximity force approximation (PFA) \cite{bordag}. The
relative error introduced by this approximation is of order $a/R$ \cite{bordag}, and
therefore it is only a fraction of a percent in typical experimental situations.
The PFA expression for $F_{\rm ps}$ is:
\be F_{\rm ps}= -\frac{k_B T R}{2\,\pi} \sum_{l \ge 0}{\,'}
\int d^2{\bf k_{\perp}} \!\!\!\!\!\!\!\!
\sum_{\alpha={\rm TE,TM}}  \!\!\!\!\log \left(1-{r_{\alpha}^{(1)} r_{\alpha}^{(2)}{e^{-2 a
q_l}}}
\right),\label{PFA} \ee
where a plus sign for the force again corresponds to attraction.  Analogously to what
we did in the plane-parallel case, we consider two distinct setups: in the first one,
both the plate and the sphere are made of Nb, while in the second setup the sphere is made
  of Nb and the plate of Au (or viceversa).  For either setup, we consider then the  change in
  the sphere-plate force
  \be{\Delta F}_{\rm ps}(a \vert T_2,s;T_1,s)\equiv F_{\rm ps}(a \vert T_2,s)-
  F_{\rm ps}(a\vert T_1,s)\,,\ee
that occurs when the equilibrium temperature of the system is
increased from from $T_1$ to   $T_2$   (the notation adopted here
is analogous to that used in the plane parallel case). By
repeating the reasonings that led us to write Eq.(\ref{keysplit}),
we can obtain the following expression for ${\Delta F}_{\rm ps}(a
\vert T_2,s;T_1,s)$
\begin{widetext}
\be {\Delta F}_{\rm ps}(a \,\vert T_2,s; T_1,s) = [{\Delta F}_{\rm
(ps) 0}^{(\rm TE)} (a \,\vert T_2,s; T_1,s)- {\Delta F}_{\rm (ps)
0}^{(\rm TE)}(a \,\vert T_2,n; T_1,n)]+ {\Delta F_{\rm ps}}(a
\,\vert T_2,n; T_1,n)\,,\label{keysplitsp} \ee
\end{widetext}
where again the meaning of the symbols is obvious. By similar
steps that led us to write Eq.(\ref{plasmaeff}), one can show
that, both for the Nb-Nb and the Nb-Au setups, the plasma
prediction for the change of  force in the sphere-plate geometry
is: \be {\Delta F}_{\rm ps}(a \,\vert T_2,s; T_1,s)\vert_{\rm pl}
= {\Delta F}_{\rm ps}(a \,\vert T_2,n; T_1,n)\vert_{\rm pl}\;. \ee
Once the small difference between the plasma frequencies of Nb and
Au is neglected, we can use the explicit expressions provided in
Ref.\cite{chen} to evaluate ${\Delta F}_{\rm ps}(a \,\vert T_2,n;
T_1,n)\vert_{\rm pl}$ (recall that our sign convention for the
Casimir  force is opposite to that of \cite{chen}):
\begin{eqnarray}
  {\Delta F}_{\rm ps}(a \,\vert T_2,n; T_1,n)\vert_{\rm pl}=\;\;\;\; && \nonumber \\
   R \Delta^{(1)}F_{ps}(T_1,T_2)\,\Delta^{(2)}F_{ps}(a,T_1,T_2),& &
\end{eqnarray}
where
\be
\Delta^{(1)}F_{ps}(T_1,T_2)=\frac{\zeta(3) k_B^3}{\hbar^2 c^2}(T_2-T_1)(T_1^2+T_2^2),
\ee
and
\begin{eqnarray}
  \!\!\!\Delta^{(2)}F_{ps}(a,T_1,T_2) &=& \left(1+\frac{T_1 T_2}{T_1^2+T_2^2}\right)
  \left(1+2 \frac{\delta}{a}\right) \nonumber\\
  &-& \frac{\pi^3}{45 \zeta(3)}\frac{T_1+T_2}{T_{\rm eff}}\left(1+4\frac{\delta}{a}\right).
\end{eqnarray}
Analogously to what we found in the plane-parallel case, the
effect predicted by the plasma prescription is unmeasurably small.
For example, for a sphere of radius $R=200$ $\mu$m, at a distance
$a=150$ nm from a plane, for $T_1=5$ K and $T_2 \simeq T_c$ the
above formulae give a change of Casimir  force of about $5.3\times
10^{-19}$ N.

We consider now the Drude prescription. Repeating the steps that
led to Eq.(\ref{DrudeNbAu}), we obtain for the Nb-Au setup the
Equation: \be {\Delta F}_{\rm ps}(a \,\vert T_2,s;
T_1,s)\vert_{\rm Dr}^{\rm Nb-Au} = {\Delta F}_{\rm ps}(a \,\vert
T_2,n; T_1,n)\vert_{\rm Dr}^{\rm Nb-Au}\,.\label{DrudeNbAu} \ee An
explicit formula for ${\Delta F}_{\rm ps}(a \,\vert T_2,n;
T_1,n)\vert_{\rm Dr}^{\rm Nb-Au}$ is given in Eq.(10) of
Ref.\cite{chen}:
\begin{eqnarray}
  {\Delta F}_{\rm ps}(a \,\vert T_2,n; T_1,n)\vert_{\rm Dr}^{\rm Nb-Au} =&&
 R \Delta^{(1)}F_{ps}\Delta^{(2)}F_{ps} \nonumber \\
   - \frac{R k_B \zeta(3)}{8 a^2}(T_2-T_1) &&\!\!\left(1- 4\frac{\delta}{a}+
   12\frac{\delta^2}{a^2}\right).
\end{eqnarray}
The first term on the right hand side of the above Equation is
completely negligible with respect to the second term. We
therefore see that the Drude model predicts a decrease of Casimir
force proportional to $T_2-T_1$. We repeat that this effect has
nothing to do with  superconductivity, and it is again analogous
to what was found in Ref.(\cite{chen}) (see comments following
Eq.(\ref{DrNbAu})).
\begin{figure}
\includegraphics{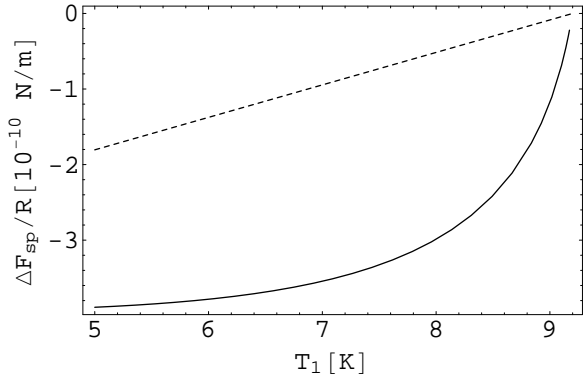}
\caption{\label{fig3}
Plots of
${\Delta F}_{\rm sp}({\rm Nb}-{\rm Nb})\vert_{\rm Dr}/R$ (solid line) and
 ${\Delta F}_{\rm sp}({\rm Nb}-{\rm Au})\vert_{\rm Dr}/R$ (dashed line), in units of $10^{-10}$ N/m,
 versus $T_1$ (in K), for $T_2 \simeq T_c$ and for fixed sphere-plate separation $a=150$ nm.}
\end{figure}

\begin{figure}
\includegraphics{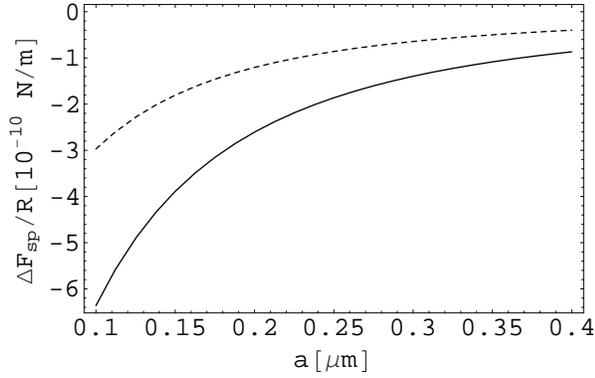}
\caption{\label{fig4} Plot of
${\Delta F}_{\rm sp}({\rm Nb}-{\rm Nb})\vert_{\rm Dr}/R$ (solid line) and
 ${\Delta F}_{\rm sp}({\rm Nb}-{\rm Au})\vert_{\rm Dr}/R$ (dashed line),
 in units of $10^{-10}$ N/m,
 versus sphere-plate separation $a$ (in microns), for $T_1=5$ K and $T_2 \simeq T_c$.}
\end{figure}

It remains to discuss the Nb-Nb setup.  By proceeding in the same way as we did to
derive  Eq.(\ref{DrNbNb}), one can prove the following Equation
\begin{widetext}
$$ {\Delta F}_{\rm ps}(a \,\vert T_2,s; T_1,s)\vert_{\rm Dr}^{\rm
Nb-Nb}=[F_{\rm (ps)0}^{\rm (TE)} (a,T_2;c/\lambda_L)-F_{\rm
(ps)0}^{\rm (TE)} (a,T_1;c/\lambda_L)]$$ \be + \,R\,
\Delta^{(1)}F_{ps}\,\Delta^{(2)}F_{ps}
  -  \frac{R k_B \zeta(3)}{8 a^2}(T_2-T_1) \left(1-4 \frac{\delta}{a}+
   12\frac{\delta^2}{a^2}\right),\label{DrNbNbsp}
\ee
\end{widetext}
where $F_{\rm (ps)0}^{\rm (TE)}(a,T;c/\lambda_L)$ denotes the TE
zero mode contribution to the sphere-plate force
 (the l=0 term for TE polarization in Eq.(\ref{PFA})), as it results
after we use  our Drude prescription for the reflection
coefficient of the superconductor, Eq.(\ref{drsup}). In Fig.3, we
plot ${\Delta F}_{\rm ps}({\rm Nb}-{\rm Nb})\vert_{\rm Dr}/R$
(solid line), and ${\Delta F}_{\rm ps}({\rm Nb}-{\rm
Au})\vert_{\rm Dr}/R$ (dashed line), in units of $10^{-10}$ N/m,
as a function of $T_1$ (in K), for  $T_2 \simeq T_c$ and for a
plate-sphere separation $a=150$ nm. In Fig.4 the same quantities
are plotted versus plate-sphere separation $a$ (in microns), for
fixed temperatures $T_1=5$ K and $T_2 \simeq T_c$. Note that the
changes of force are always negative, indicating that the Casimir
force decreases as the temperature of the superconducting system
is increased from $T_1$ to  $T_c$. As we see from Fig.3, for a
sphere-plate separation of 150 nm and for a sphere of radius
$R=200$ $\mu$m, the Drude model predicts for the Nb-Nb setup a
change in the force around $-0.8 \times 10^{-13}$ N, which
 is over five orders of magnitude larger than the plasma model prediction!

\section{Conclusions and discussion}

One of the most important unresolved theoretical problems in the
theory of dispersion forces is that of determining the thermal
correction to the Casimir force between two conductors. The source
of the difficulties stems from the ambiguity in the correct value
for the ${\rm TE}$ zero-mode contribution to the force,
corresponding to quasi-static magnetic fields. In this paper, we
have proposed a new Casimir experiment with superconducting
cavities, specifically devised to probe the ${\rm TE}$ zero-mode.
The proposal involves measuring the {\it change}  of  Casimir
force  that occurs in a cavity with one or two superconducting
plates, as the temperature of the device is increased towards the
normal state. The interest of performing difference force
measurements to
 probe the features of the thermal correction to the Casimir force has already
been stressed in Ref.\cite{chen}, which considered measuring the difference in the
Casimir force at two different temperatures in cavities made of ordinary metals at
room temperature. Our proposal is very much in the same spirit as Ref.\cite{chen},
with the significant difference that the change in the Casimir force determined by a
 small change of
temperature is much enhanced (in the Drude case) by
superconductivity of the plates. The  need of smaller temperature
changes, than those required in \cite{chen},
 might make it easier to achieve them in an experiment.

It is perhaps the case to comment further on the possible
advantages and drawbacks that are implicit in the proposed scheme.
A potential advantage is that, being a {\it difference force}
measurement, it  might  be possible to achieve a better
sensitivity than in an absolute force measurement. This of course
holds true only provided that  systematic errors  do not change
too much when the temperature of the  apparatus is varied.
Importantly enough, this is the case with regards to the
theoretical uncertainty arising from insufficient knowledge of the
optical data of the plates, that was estimated in Ref. \cite{piro}
to be easily as large as five percent in absolute force
measurements. This uncertainty is completely irrelevant in our
scheme, because as we showed in Sec.III, the magnitude of the
effect is determined solely by the ${\rm TE}$ zero mode.  More
delicate is the case of the systematic
 errors arising from residual electrostatic  attractions between the plates \cite{patch},
 which constitute
a concern in all Casimir experiments.  If their magnitude is
temperature dependent in the superconducting state,  it might be
necessary to perform   electrostatic calibrations of the apparatus
at the considered temperatures.

One of the main experimental issues to address is to find a good
mean of  varying  the  temperature  of the plates. Perhaps, a
convenient way to  do this is by illuminating the plates with
short laser pulses, as suggested in Ref.\cite{chen}. Very likely,
this
 method of heating the plates will reduce to a minimum unwanted thermal expansions/contractions
of parts of the apparatus that would inevitably  result, were we to warm up the
whole system instead. Such expansions would clearly easily alter the separations
 of the plates by a few nanometers, thus rendering much more difficult a comparison of the Casimir
 force between the two temperatures.

Finally, we would like to comment on the prospects of the
experiment described in this paper of being actually capable of
discriminating between the contradictory theoretical approaches to
the description of the thermal Casimir force between real
materials. According to our computations, the two approaches
predict strikingly different magnitudes for the temperatures
changes of Casimir force in the superconducting state, the plasma
approach predictions being always  five or six orders of magnitude
smaller than those of the Drude approach. While the changes of
force resulting from the plasma approach are unmeasurably small,
the effect predicted by the Drude approach is in fact large enough
for an experimental detection to be hopeful. For example, in the
sphere-plate geometry, which is the most widely used geometry in
current experiments, if  both the sphere and the plate are made of
Nb, for a sphere radius of 200 $\mu$m, and a sphere-plate
separation of 150 nm the Drude  approach predicts a drop in the
Casimir force of $0.8 \times 10^{-13}$ N, as the system  is heated
by two-three degrees. Force oscillations slightly larger than this
one have been demonstrated to be measurable by means of an atomic
force microscope, with an error about 20 percent at 95 percent
confidence level, in a recent measurement of the so-called lateral
Casimir force \cite{chen2}. If the same accuracy can be achieved
at cryogenic temperatures, there are good chances that the
experiment proposed here might actually be able to discriminate
the alternative approaches to the description of the thermal
Casimir force between real metals.

\section{Ackowledgements}

The author thanks R. Decca, G.L. Klimchitskaya, V. Mostepanenko  for their
critical reading of the manuscript, and T. Ricciardi for her invaluable support.
The authors acknowledges also the kind hospitality of the Kavli
Institute for Theoretical Physics in Santa Barbara, where this work was completed.
This research was supported in part by the National Science Foundation under Grant No.
NSF PHY05-51164.

\end{document}